\documentclass[12pt]{article}
\usepackage[top = 2 cm, bottom = 2 cm, left = 2 cm, right = 1.5 cm]{geometry}
\usepackage{authblk, cite, color, amssymb, amsmath}
\usepackage[hypertex, colorlinks = true, linkcolor = blue, citecolor = red]{hyperref}
\usepackage[dvipsnames]{xcolor}

\begin{document}

\title{The energy meaning of Boltzmann's constant}

\author{{\bf Merab Gogberashvili$^{1,2}$}}
\affil{\small $^1$ Javakhishvili Tbilisi State University, 3 Chavchavadze Avenue, Tbilisi 0179, Georgia \authorcr
$^2$ Andronikashvili Institute of Physics, 6 Tamarashvili Street, Tbilisi 0177, Georgia \authorcr}

\maketitle

\begin{abstract}
The atomic scale is the only relevant thermodynamic scale in our universe, since quantum properties restrict classical considerations of subatomic physics and disappear for larger scales. Then the characteristic energy that dictates the value of the unit of temperature can be the classical thermal energy defined for simplest atoms. It is shown that vibrational frequency of a classical model hydrogen atom, of the radius of its Rydberg wavelength, is in far-infrared range and from its quantum of energy one can obtain the value of Boltzmann's constant that serves as the measure of the absolute temperature in kelvins.

\end{abstract}


Temperature $T$ is a basic physical quantity characterizing the states of macroscopic bodies enabling one to describe thermal equilibrium between systems in thermal contact on the basis of the zeroth law of thermodynamics \cite{Termo}. However, this law does not say anything about the origin of $T$, nor fix its unit, which is the Kelvin ($K$) in the international system of units (SI).

The ambiguity in introducing the temperature by the zeroth law makes it possible to choose another unit of measurement, more appropriate to the physical mechanism. Since $T$ always appear in basic physical laws together with one of the ingredients of the set of fundamental physical constants -- Boltzmann's constant
\begin{equation} \label{k}
k = 1.38 \times 10^{-23}~\frac JK~,
\end{equation}
one can take as the temperature the combination \cite{Atkins, Ka-Ko}
\begin{equation} \label{theta}
\theta = kT ~.
\end{equation}
The unit of measurement for $\theta$ is an energy measurement unit, the Joule ($J$) in SI. The quantity $\theta$ appears in distribution functions of equilibrium statistical mechanics in the classical and quantum cases and in the relation of temperature to the mean kinetic energy of molecules per degree of freedom,
\begin{equation} \label{Ekin}
\langle E_{\rm kin} \rangle= \frac 12 \,\theta~.
\end{equation}
However, in the classical thermodynamic formula of increment of heat transfer,
\begin{equation}
dQ = T dS~,
\end{equation}
Boltzmann's constant $k$ is considered as the part of the entropy $S$. There is no single universally accepted definition of entropy in statistical physics. Then, as in (\ref{theta}), it is natural to relate $k$ to the temperature $T$ (not to the thermodynamic entropy $S$) and to introduce the dimensionless quantity
\begin{equation}
H = \frac Sk~,
\end{equation}
which agrees with the entropy definition in Shannon’s information theory. Shannon’s information entropy $H$ is determined for any system (not necessarily the thermodynamic one) characterized by a certain set of random variables \cite{Inform}.

The above analysis that thermodynamic entropy is naturally to take as a dimensionless quantity shows that Boltzmann’s constant $k$ may be considered as a unit of the absolute temperature $T$ and try to obtain its value considering characteristic heat transfer processes that correspond to the unit of the energy measure of temperature (\ref{theta}).

In standard thermodynamics the world is divided into microscopic and macroscopic objects and both the existence of and an asymmetry between different scales has to be postulated. Without such an asymmetry, and if there were many more macroscopic than microscopic objects, thermodynamic laws would not hold.

The absolute temperature $T$ is a classical quantity that relates with the mechanical velocities of atoms and molecules and considered to be zero in absence of classical motions. Although temperature could be identified not only with the movement of atoms and molecules, but also with that of macroscopic objects, or some hypothetical classical microscopical particles that are much smaller than atoms. However, the scaled-up thermodynamics do not hold, there is no lower or higher scales. It seems that the atomic scale is the only relevant thermodynamic scale in our universe, there are no particles smaller than atoms which can be used to build solid structures. The reason is quantum properties of nature, which restrict classical considerations of subatomic physics and disappear for large scales, but still enable creation of ensembles of particles that can be attributed by trajectories and velocities.

It is known that one can consider laws of thermodynamics in other contexts, e.g. Hawking \cite{Hawking} and Unruh \cite{Unruh} temperatures. However, these models just employ analogies of thermodynamic behaviors and for other than atomic-molecular systems it is not easy to justify use of the absolute temperature and Boltzmann’s constant.

In our opinion the characteristic scale that dictates the value of the unit of temperature can be the Rydberg wavelength of the hydrogen atom,
\begin{equation} \label{r}
r \simeq 10^{-7}~ m~.
\end{equation}
The size of the real hydrogen atom is $\sim 10^{-10}~m$, but atom is a quantum object and cannot be described by classical physics. Thermodynamic definition of the absolute temperature by classical motions of particles (\ref{Ekin}) ignores subatomic vibrations at the scales smaller than (\ref{r}).

Thermodynamically hydrogen atom can be treated as a sphere of radius (\ref{r}) that vibrate under the mutual field of force between proton and electron. The vibrational energy states of quantum harmonic oscillator have energies given by
\begin{equation} \label{E}
E_n = \left(n + \frac 12\right)\hbar \omega ~, \qquad (n = 0, 1, 2, ...)
\end{equation}
where
\begin{equation}
\hbar = 1.05 \times 10^{-34}~ J\cdot s
\end{equation}
is reduced Planck's constant and
\begin{equation} \label{omega}
\omega = \sqrt{\frac \sigma \mu}
\end{equation}
is the fundamental angular frequency of oscillations of proton-electron system. In (\ref{omega}) the quantity $\sigma$ is the elastic constant of this vibration and
\begin{equation}
\mu \approx m_e \approx 10^{-30}~ kg
\end{equation}
is the reduced mass of the proton-electron system.

Zero temperature, $T = 0$, corresponds to the vacuum energy, when $n = 0$ in (\ref{E}). Then the characterized thermal energy of the model hydrogen atom vibrations (three-dimensional oscillator), which can be relevant to define the unit of $\theta$, can be
\begin{equation} \label{theta-0}
3\theta_0 = \hbar \omega ~,
\end{equation}
the difference in energy when the vibrational quantum number $n$ in (\ref{E}) changes by 1.

To estimate the angular frequency of oscillations (\ref{omega}) note that the binding of model hydrogen atom is produced by the Coulomb action of proton and electron in their stable states. Then the potential energy of the oscillator of the size $r$ is of the order of the electric energy of a charged sphere,
\begin{equation}
\sigma r^2 \simeq \frac 35 \frac {k_c e^2}{r}~,
\end{equation}
where
\begin{equation}
k_c = 8.99 \times 10^9~ \frac {kg\cdot m^3}{s^2\cdot C^2}~, \qquad e = 1.60 \times 10^{-19}~ C
\end{equation}
are Coulomb's constant and the electric charge, respectively. Then
\begin{equation}
\sigma \simeq \frac {3k_c e^2}{5r^3}
\end{equation}
and the frequency of oscillations (\ref{omega}) appears to be in far-Infrared range:
\begin{equation} \label{}
\omega = \sqrt{\frac {3k_c e^2}{5 m_er^3}} \approx 4 \times 10^{11} ~\frac {1}{s}~,
\end{equation}
which is about rotational frequencies of diatomic molecules and is very different from the lowest frequencies of the hydrogens Lyman series $\sim 10^{16}~s^{-1}$. The reason for this mismatch is that for thermodynamic considerations we use not the real size of a hydrogen atom, but the Rydberg scale (\ref{r}), for which the atom can be considered as a quasi-classical oscillator having the energy (\ref{E}).

From (\ref{theta-0}) for the unit thermal energy of a model hydrogen atom we find
\begin{equation} \label{theta-0=}
\theta_0 = kT_0 = \frac 13 \, \hbar \omega \approx 1.4 \times 10^{-23} ~J~,
\end{equation}
which explains the value of Boltzmann's constant (\ref{k}) that serves as the measure of the unit of the absolute temperature, $T_0 = 1~K$. The value (\ref{theta-0=}) also reveals the characteristic number of particles (Avogadro number) one needs in thermodynamic considerations.

To conclude, note that classical thermodynamics cannot be extrapolated neither for subatomic physics, nor for large scales where quantum properties of nature does not hold good. Then the characteristic energy that dictates the value of the unit of temperature can be the classical thermal energy defined for simplest atoms. In this paper it is shown that vibrational frequency of a model hydrogen atom of the radius of its Rydberg wavelength is in far-infrared range and the numerical value of Boltzmann's constant, which serves as the measure of the absolute temperature in kelvins, is obtained from the value of its quantum of energy.

\section*{Acknowledgements}

This work was supported by Shota Rustaveli National Science Foundation of Georgia (SRNSFG) through the grant DI-18-335.



\begin{thebibliography}{4}

\bibitem{Termo} R.~Fitzpatrick,
{\it Thermodynamics and Statistical Mechanics}
(World Scientific, Singapore 2020)
\url{http://farside.ph.utexas.edu/teaching/sm1/Thermal.pdf}.

\bibitem{Atkins} P.~Atkins,
{\it Four Laws that Drive the Universe}
(Oxford University Press, Oxford 2007).

\bibitem{Ka-Ko} M.~Kalinin and S.~Kononogov,
"Boltzmann’s constant, the energy meaning of temperature and thermodynamic irreversibility,"
Meas. Tech. {\bf 48} (2005) 632,
doi: 10.1007/s11018-005-0195-9.

\bibitem{Inform} T.~M.~Cover and J.~A.~Thomas,
{\it Elements of Information Theory}
(Wiley, New York 2001).

\bibitem{Hawking} S.~W.~Hawking,
"Black hole explosions?"
Nature {\bf 248} (1974) 30,
doi:10.1038/248030a0.

\bibitem{Unruh} W.~G.~Unruh,
"Notes on black-hole evaporation,"
Phys. Rev. D {\bf 14} (1976) 870,
doi: 10.1103/PhysRevD.14.870.

\end{thebibliography}
\end{document}